\documentclass[10pt,aps,prd,twocolumn, floatfix,nofootinbib,showpacs]{revtex4}
\usepackage{graphicx}
\usepackage{color}
\begin{document}



\title{Higgs boson masses in an extension of the MSSM with vector-like quarks}

\renewcommand{\thefootnote}{\fnsymbol{footnote}}

\author{
S. W. Ham$^{(1)}$\footnote{s.w.ham@hotmail.com},
Seong-A Shim$^{(2)}$\footnote{shims@sungshin.ac.kr},
C. M. Kim$^{(3)}$\footnote{kcmh@konkuk.ac.kr},
and S. K. Oh$^{(3)}$\footnote{sunkun@konkuk.ac.kr}
 }
\affiliation{(1) School of Physics, KIAS, Seoul 130-722, Korea \\
(2) Department of Mathematics, Sungshin Women's University, Seoul 136-742 \\
(3) Department of Physics, Konkuk University, Seoul 143-701 }

\renewcommand{\thefootnote}{\arabic{footnote}}

\begin{abstract}
We study the Higgs sector of the minimal supersymmetric standard model extended with vector-like quarks,
at the one-loop level.
The radiative corrections to the tree-level masses of the scalar Higgs bosons are calculated by including
the contributions from the loops of top quark, vector-like quarks, and their scalar superpartners, for a reasonable parameter region.
We find that the mass of the lightest scalar Higgs boson at the one-loop level should be larger than
85 GeV, if we take into account the negative experimental result for the Higgs search at LEP2.
As the radiative corrections are calculated in some detail, we also find that the mass of the lightest scalar Higgs
boson at the one-loop level is bounded from above at 280 GeV,
This upper bound is increased from a previous result.
It may provide a wider possibility for the future collider experiments to discover the lightest scalar Higgs boson of this model.
\end{abstract}


\maketitle

\section{Introduction}

Although the minimal supersymmetric standard model (MSSM) [1-5] is the simplest version among
the supersymmetric extensions of the Standard Model (SM), with just two Higgs doublets,
there are some arguments that it should be extended [6].
One of the motivations to extend the MSSM may be the fine-tuning problem,
which states that the radiative corrections to the tree-level mass of the lightest scalar Higgs
boson in the MSSM should be large enough to be consistent with the  experimental lower bound on
the Higgs boson mass at LEP2, around 114 GeV.
If the radiative corrections are not large enough, the lightest scalar Higgs boson in the MSSM might face
the difficulty of incompatibility with experiments.

In the next-to-MSSM [7-10], where an additional Higgs singlet is introduced to the MSSM,
the fine-tuning problem may be alleviated by the contributions from the Higgs singlet.
The presence of the Higgs singlet increases the upper bound on the tree-level mass of the lightest scalar
Higgs boson in the next-to-MSSM up to about 115 GeV, and thus the radiative corrections need not be
large enough to be consistent with LEP2 data [11-38].

The burden of the fine tuning in the MSSM might also be reduced by adding extra quarks, such as
the sequential quarks or the vector-like quarks.
If the additional quarks are those of the fourth generation, with masses at about 400 GeV,
it has been shown that the mass of the lightest scalar Higgs boson in the MSSM might increase up to 400 GeV [39].
If the additional quarks are vector-like quarks, as proposed about two decades ago by Moroi and Okada [40,41],
the fine tuning problem might be solved by the radiative corrections due to
the contributions from the vector-like quarks and the vector-like scalar quarks.
The vector-like quarks and their superpartners increase the mass of the lightest scalar Higgs boson
well above the LEP2 bound, without requiring a large SUSY breaking scale as 2 TeV [40-46].

Recently, in the extension of the MSSM with vector-like quarks, Martin has investigated
the one-loop corrections to the mass of the lightest scalar Higgs boson [46].
He has shown that the radiative corrections due to the vector-like quarks and vector-like scalar quarks
may be as large as about 55 GeV, if the mass of the lightest scalar Higgs boson before correction is 110 GeV.
His result has been obtained by  neglecting some soft SUSY breaking parameters to simplify
the $4 \times 4$ mass matrix of the vector-like scalar quarks into two $2 \times 2$ submatrices,
and assuming that the weak eigenstates and the mass eigenstates of the vector-like quarks are identical.

In this article, we show that the one-loop mass of the lightest scalar Higgs boson may be even as
large as 280 GeV if we carry out the calculations without the above simplifications assumed in Ref. [46],
by taking into account the contributions from top quark and vector-like quarks and their scalar superpartners
at the one-loop level.
The masses of the vector-like quarks are assumed to lie in the range of 300 GeV to 550 GeV, and
the masses of the vector-like scalar quarks in the range of 200 GeV to 910 GeV.
The lower bound on the mass of the lightest scalar Higgs boson in this model is determined
such that it should not have been discovered at LEP2 with a mass smaller than its lower bound.
The negative experimental result for the Higgs search at LEP2 set the
lower bound on the mass of the lightest scalar Higgs boson as 85 GeV.
We may note that, since the mass of the lightest scalar Higgs boson can be sufficiently large,
there are wide regions in the parameter space of the extensions of the MSSM with vector-like quarks
which shall be examined by the Higgs search in future high-energy collider experiments.

\section{Higgs Bosons}

Martin have investigated several versions of extensions of the MSSM with the vector-like matters in Ref. [46].
To be specific, we study the QUE model among them, where three vector-like chiral fields
are introduced to the MSSM.
These vector-like fields transform under $SU(3)_C \times SU(2)_L \times U(1)$ as
\begin{eqnarray}
& & {\cal Q} = ({\bf 3}, {\bf 2}, 1/6),  \quad {\bar {\cal Q}} = ({\bar {\bf 3}}, {\bf 2}, - 1/6), \cr
& & {\cal U} = ({\bf 3}, {\bf 1}, 2/3),  \quad {\bar {\cal U}} = ({\bar {\bf 3}}, {\bf 1}, - 2/3), \cr
& & {\cal E} = ({\bf 1}, {\bf 1}, - 1),  \quad {\bar {\cal E}} = ({\bar {\bf 1}}, {\bf 1}, 1),
\end{eqnarray}
where ${\cal Q}$, ${\bar {\cal Q}}$, ${\cal U}$, and ${\bar {\cal U}}$ are the vector-like quark fields,
while ${\cal E}$ and ${\bar {\cal E}}$ are the vector-like lepton fields.
In terms of physical particles, this model has, in addition to the SM quarks and leptons,
$V_1$ and $V_2$, quarks with charge $+ 2/3$,  $B'$, a quark with charge $- 1/3$, and a charged lepton $L'$,
together with their scalar partners ${\tilde V}_i$ ($i = 1,2,3,4$), ${\tilde B}_i$ ($i = 1,2$),
and ${\tilde L}_i$ ($i = 1,2$).

The superpotential of our model may be written as
\begin{eqnarray}
{\cal W} & = & {\cal W}_{\rm MSSM} + M_Q {\cal Q} {\bar {\cal Q}} + M_U {\cal U} {\bar {\cal U}}
+ M_E {\cal E} {\bar {\cal E}} \cr
& &\mbox{}  + k {\cal H}_u {\cal Q} {\bar {\cal U}} - h {\cal H}_d {\bar {\cal Q}} {\cal U} \ ,
\end{eqnarray}
where $M_Q$, $M_U$, and $M_E$ are the soft SUSY breaking masses of the vector-like sector,
and $k$ and $h$ are Yukawa coupling coefficients
to the Higgs fields $H_u$ and $H_d$, with the weak hypercharge $+ 1/2$ and $- 1/2$, respectively.

The tree-level Higgs potential of our model is the same as the Higgs potential of the MSSM at the tree level.
It is given as
\begin{eqnarray}
V_0 & = & m_u^2 |H_u|^2 + m_d^2 |H_d|^2 - \bigg (m_{ud}^2 H_u H_d + {\rm H.c.} \bigg )   \cr
&  &\mbox{} + {1 \over 8} ({g_1}^2 + g_2^2) \bigg ( |H_u|^2 - |H_d|^2  \bigg )^2    \  ,
\end{eqnarray}
where $m_u$, $m_d$, and $m_{ud}$ are the mass parameters,
$H_d^T = (H_d^0, H_d^-)$ and $H_u^T = (H_u^+, H_u^0)$ are two Higgs doublets,
and $g_1$ and $g_2$ are respectively the gauge coupling coefficients for $U(1)$ and $SU(2)$.
There are two neutral scalar Higgs bosons and a neutral pseudoscalar Higgs bosons in the MSSM.
The vacuum expectation values (VEVs) developed by two neutral scalar Higgs bosons are
denoted as $v_d = \langle H_d \rangle$ and $v_u = \langle H_u \rangle$,
and their ratio is repesented by $\tan \beta  = v_u / v_d$.
In this article, we would not consider the CP mixing between the scalar and pseudoscalar Higgs bosons.
Thus, the vacuum expectation values of the neutral Higgs fields as well as
the parameters of the Higgs potential are all real.

Two mass parameters $m_d$ and $m_u$ can be eliminated by the minimum equations
with respect to the scalar Higgs fields.
Thus, just two paramaters, $m_{ud}$ and $\tan \beta$, describe the tree-level masses of the scalar Higgs bosons.
They are given by the eigenvalues of the tree-level $2 \times 2$ mass matrix, $M^0$, for the scalar Higgs bosons.
The matrix elements of $M^0$ are given explicitly as
\begin{eqnarray}
M_{11}^0 & = & m_Z^2 \cos^2 \beta + m_{ud}^2 \tan \beta     \ , \cr
M_{22}^0 & = & m_Z^2 \sin^2 \beta + m_{ud}^2 \cot \beta    \ , \cr
M_{12}^0 & = &\mbox{} - m_Z^2 \cos \beta \sin \beta - m_{ud}^2   \ ,
\end{eqnarray}
where $m_Z^2 = (g_1^2 + g_2^2) v^2/2$ is the squared mass of $Z$ boson,
with $v = \sqrt{v_d^2 + v_u^2} = 175$ GeV.
Note that the tree-level mass of the lightest scalar Higgs boson is smaller than $Z$ boson mass.

In order to study the radiative corrections to the tree-level Higgs boson mass,
we evaluate the Higgs potential at the one-loop level.
The one-loop effective potential is given by the effective pottential approximation as [47]
\begin{equation}
    V_1 = \sum_{l} {n_l {\cal M}_l^4 \over 64 \pi^2}
    \left [ \log {{\cal M}_l^2 \over \Lambda^2} - {3 \over 2} \right ]  \ ,
\end{equation}
where $\Lambda$ is the renormalization scale in the modified minimal subtraction scheme,
${\cal M}_l$ are the field-dependent masses of quarks and scalar quarks,
and $n_l$ are the degrees of freedom arising from color, charge, and spin factors of the particles in the loops,
hence, $n_l = -12$ for quarks and $n_l =6$ for scalar quarks.
We take into account top quark, vector-like quarks and their scalar superpartners,
as their contributions are significant for the mass of the lightest scalar Higgs boson
at the one-loop level in our analysis.
The full Higgs potential at the one-loop level is thus
\begin{equation}
    V = V_0 + V_1 \ .
\end{equation}

The masses of the scalar top quarks after electroweak symmetry breaking may be expressed as
\begin{equation}
m_{{\tilde t}_1, {\tilde t}_2}^2 =  {(m_Q^2 + m_T^2) \over 2} + m_t^2 \mp \sqrt{X_t} \ ,
\end{equation}
with
\begin{equation}
X_t = \bigg [ {m_Q^2 - m_T^2 \over 2} \bigg ]^2 + m_t^2 \bigg ( A_t - \mu \cot \beta \bigg)^2 \ ,
\end{equation}
where $m_Q$ and $m_T$ are soft SUSY breaking masses,
$m_t = h_t v_u $ is the mass of top quark, with $h_t$ being the Yukawa coupling coefficient for top quark,
$A_t$ is the trilinear SUSY breaking mass parameter,
and $\mu$ is a parameter with mass dimension.
Note that $X_t$ represents the mixing between the left-handed and the right-handed scalar top quarks.

Next, let us study the vector-like sector of our model.
We adapt the notation of Re. [46].
The mass matrix for the vector-like quarks at the tree level is given by
\begin{equation}
m_F^2 =
\left(\begin{array}{cc}
{\cal M}_F {\cal M}_F^{\dagger} & 0  \\
0 & {\cal M}_F^{\dagger}{\cal M}_F   \
\end{array} \right)  \ ,
\end{equation}
with
\begin{equation}
{\cal M}_F =
\left(\begin{array}{cc}
M_Q & k v_u  \\
h v_d & M_U  \
\end{array} \right)  \ ,
\end{equation}
The squared masses of the two vector-like quarks, $m^2_{V_i}$ ($i = 1,2$),
are given by the doubly degenerate eigenvalues of $m_F^2$.

\begin{widetext}
The mass matrix of the vector-like scalar quarks is given by
\begin{equation}
M_{\tilde V}^2 = m_F^2 +
\left(\begin{array}{cccc}
m_1^2 & 0 & b_{\Phi} & A_k v_u - k \mu v_d  \\
0 & m_2^2 & A_h v_d - h \mu v_u & b_{\phi}   \\
b_{\Phi} & A_h v_d - h \mu v_u & m_3^2 & 0   \\
A_k v_u - k \mu v_d & b_{\phi} & 0 & m_4^2
\end{array} \right)  \ ,
\end{equation}
\end{widetext}
where $b_{\Phi}$ and $b_{\phi}$ are the soft SUSY breaking parameters with mass-square dimension,
$m_1$, $m_2$, $m_3$ and $m_4$ are the soft SUSY breaking masses and
$A_k$ and $A_h$ are the trilinear mass parameters.
We assume that they are all real and yield real eigenvalues.
We note that, unlike Ref. [46], $M_{\tilde V}^2$ cannot decomposed into two $2 \times 2$ submatrices,
since we keep non-zero off-diagonal elements in $m_F^2$.
Moreover, we do not neglect $b_{\Phi}$ and $b_{\phi}$.
This is the main difference from Ref. [46].

Now, we apply the above formulae for the tree-level masses of top quark, vector-like quarks, and
their scalar superpartners into the one-loop effective Higgs potential
in order to evaluate their contributions to the mass of the lightest scalar Higgs boson.
The masses of two scalar Higgs bosons at the one-loop level are given as
\begin{equation}
m_{{S_1}, {S_2}}^2 = {1 \over 2} \bigg [ {\rm Tr} \bigg (M \bigg )
\mp \sqrt{ \bigg ({\rm Tr}(M) \bigg )^2 - 4 {\rm det} \bigg (M \bigg ) } \bigg ]  \ ,
\end{equation}
where $S_1$ is the lighter scalar Higgs boson as we define $m_{S_1} < m_{S_2}$,
and $M$ is the $2\times 2$ squared mass matrix for the scalar Higgs bosons at the one-loop level.

It is convenient to decompose $M$ as
\begin{equation}
M = M^0 + M^t + 2 M^V + M^{\tilde V} \ ,
\end{equation}
where $M^0$ is the tree-level mass matrix, and $M^t$, $M^V$, and $M^{\tilde V}$ respectively denotes
the radiative corrections due to the contributions from the loops of top quark and scalar top quarks,
from the loops of the vector-like quarks, and from the loops of the vector-like scalar quarks.
Note that the factor of 2 on $M^V$ shows that two vector-like quarks produce the same radiative corrections
due to the degeneracy between them.
Let us calculate each of these radiative corrections.

The radiative corrections due to the loops of top quark and scalar top quarks are given as ($i,j = 1,2$)
\begin{eqnarray}
M^t_{ij} & = & {3 W_i^t W_j^t \over 32 \pi^2 v^2}
{g(m_{{\tilde t}_1}^2, m_{{\tilde t}_2}^2) \over (m_{{\tilde t}_2}^2 - m_{{\tilde t}_1}^2)^2}
+ {3 A_i^t A_j^t \over 32 \pi^2 v^2}
\log \left ( {m_{{\tilde t}_1}^2 m_{{\tilde t}_2}^2 \over \Lambda^4 } \right ) \cr
& &\mbox{} + {3 \over 32 \pi^2 v^2} (W_i^t A_j^t + A_i^t W_j^t)
{ \log ( m_{{\tilde t}_2}^2/ m_{{\tilde t}_1}^2)  \over (m_{{\tilde t}_2}^2 - m_{{\tilde t}_1}^2)} \cr
& &\mbox{} + D_{ij}^t  \  ,
\end{eqnarray}
where
\begin{equation}
 g(m_x^2,m_y^2) = {m_y^2 + m_x^2 \over m_x^2 - m_y^2} \log {m_y^2 \over m_x^2} + 2 \ ,
\end{equation}
\begin{eqnarray}
D^t & = &\mbox{} - {3 \over 16 \pi^2 v^2}
\left( {m_t^2 \mu A_t \cos \varphi \over \sin^3 \beta \cos \beta} \right)
f(m_{{\tilde t}_1}^2, \ m_{{\tilde t}_2}^2) \ , \cr
& & \cr
D_{11}^t & = & \sin^2 \beta D^t \ , \cr
& & \cr
D_{22}^t
& = & \cos^2 \beta D^t   \ , \cr
& & \cr
D_{12}^t
& = &\mbox{} - \cos \beta \sin \beta D^t    \ ,
\end{eqnarray}
and
\begin{eqnarray}
A_1^t & = & 0 \ , \cr
A_2^t & = & {2 m_t^2 \over \sin \beta}   \ , \cr
W_1^t & = & {2 m_t^2 \mu (\mu \cot \beta - A_t) \over \sin \beta }   \ , \cr
W_2^t & = & {2 m_t^2 A_t (A_t - \mu \cot \beta) \over \sin \beta }   \ ,
\end{eqnarray}
with
\begin{equation}
f(m_x^2, m_y^2) = {1 \over (m_y^2 - m_x^2)} \left[  m_x^2 \log {m_x^2 \over \Lambda^2} - m_y^2
\log {m_y^2 \over \Lambda^2} \right] + 1 \ .
\end{equation}

The radiative corrections due to the loops of the vector-like quarks are given as ($i,j = 1,2$)
\begin{eqnarray}
M^V_{ij} & = & {3 W_i^V W_j^V \over 32 \pi^2 v^2}
{g(m_{V_1}^2, m_{V_2}^2) \over (m_{V_2}^2 - m_{V_1}^2)^2}  \cr
& &\mbox{} + {3 A_i^V A_j^V \over 32 \pi^2 v^2}
\log \left ( {m_{V_1}^2 m_{V_2}^2 \over \Lambda^4 } \right ) \cr
& &\mbox{} + {3 \over 32 \pi^2 v^2} (W_i^V A_j^V + A_i^V W_j^V)
{ \log ( m_{V_2}^2/ m_{V_1}^2)  \over (m_{V_2}^2 - m_{V_1}^2)} \cr
& &\mbox{} + D_{ij}^v \  ,
\end{eqnarray}
where
\begin{eqnarray}
A_1^V & = & v v_d h^2 , \cr
& & \cr
A_2^V & = & v v_u k^2 , \cr
& & \cr
W_1^V & = & v v_d h^2 (4 M_Q^2 + 5 v_d^2 h^2 - v_u^2 k^2)   \ , \cr
& & \cr
W_2^V & = & v v_u k^2 (v_u^2 k^2 - v_d^2 h^2)   \ ,
\end{eqnarray}
where $m_{V_i}$ ($i=1,2$) are the masses of the vector-like quarks.

The radiative corrections due to the loops of the vector-like scalar quarks are given as ($i,j = 1,2$)
\begin{eqnarray}
M_{ij}^{{\tilde V}}  & = & \sum_{k=1}^4
{3 m_{{\tilde V}_k}^2 \over 32 \pi^2}
\left (\log {m_{{\tilde V}_k}^2 \over \Lambda^2} - 1 \right )
{\partial^2 m_{{\tilde V}_k}^2 \over \partial S_i \partial S_j} \cr
& &\mbox{}
+ \sum_{k=1}^4 {3 \over 32  \pi^2} \log{m_{{\tilde V}_k}^2 \over \Lambda^2} \left
({\partial m_{{\tilde V}_k}^2 \over \partial S_i}  \right)
\left( {\partial  m_{{\tilde V}_k}^2 \over \partial S_j} \right )  \ ,
\end{eqnarray}
where the first-order derivative $\partial m_{{\tilde V}_k}^2 / \partial S_i$ is given explicitly by
\begin{equation}
        {\partial m_{{\tilde V}_k}^2 \over \partial S_i}
        = - {\displaystyle
        A_i m_{{\tilde V}_k}^6 + B_i m_{{\tilde V}_k}^4 + C_i m_{{\tilde V}_k}^2 + D_i
        \over
        4 m_{{\tilde V}_k}^6 + 3 A m_{{\tilde V}_k}^4 + 2 B m_{{\tilde V}_k}^2 + C}
  \\ ,
\end{equation}
and the second-order derivative $\partial^2 m_{{\tilde V}_k}^2 /\partial S_i \partial S_j$ by
\begin{eqnarray}
        {\partial^2 m_{{\tilde V}_k}^2 \over \partial S_i \partial S_j}
        & = & \mbox{} -{A_{ij} m_{{\tilde V}_k}^6 + B_{ij} m_{{\tilde V}_k}^4
        + C_{ij} m_{{\tilde V}_k}^2 + D_{ij}
        \over 4 m_{{\tilde V}_k}^6 + 3 A m_{{\tilde V}_k}^4 + 2 B m_{{\tilde V}_k}^2 + C } \cr
& &\mbox{}
        + {(A_i m_{{\tilde V}_k}^6 + B_i m_{{\tilde V}_k}^4 + C_i m_{{\tilde V}_k}^2 + D_i)
        \over (4 m_{{\tilde V}_k}^6 + 3 A m_{{\tilde V}_k}^4 + 2 B m_{{\tilde V}_k}^2 + C)^2 } \cr
& &\mbox{}
        \times (3 A_j m_{{\tilde V}_k}^4 + 2 B_j m_{{\tilde V}_k}^2 + C_j) \cr
& &\mbox{}
        + {(A_j m_{{\tilde V}_k}^6 + B_j m_{{\tilde V}_k}^4 + C_j m_{{\tilde V}_k}^2 + D_j)
        \over (4 m_{{\tilde V}_k}^6 + 3 A m_{{\tilde V}_k}^4 + 2 B m_{{\tilde V}_k}^2 + C)^2 } \cr
& &\mbox{}
        \times (3 A_i m_{{\tilde V}_k}^4 + 2 B_i m_{{\tilde V}_k}^2 + C_i)  \cr
& &\mbox{}
        - {(12 m_{{\tilde V}_k}^4 + 6 A m_{{\tilde v}_k}^2 + 2 B)
        \over (4 m_{{\tilde V}_k}^6 + 3 A m_{{\tilde V}_k}^4 + 2 B m_{{\tilde v}_k}^2 + C)^3 } \cr
& &\mbox{}
        \times
        (A_i m_{{\tilde V}_k}^6 + B_i m_{{\tilde V}_k}^4 + C_i m_{{\tilde V}_k}^2 + D_i) \cr
& &\mbox{}
        \times
        (A_j m_{{\tilde V}_k}^6 + B_j m_{{\tilde V}_k}^4 + C_j m_{{\tilde V}_k}^2 + D_j)
         \ .
\end{eqnarray}
The explicit expressions for the various coefficients in the above formulae are
given in the Appendices: $A$, $B$, and $C$ in Appendix I,
$A_i$, $B_i$, $C_i$, and $D_i$ $(i = 1, 2)$ in Appendix II
and $A_{ij}$, $B_{ij}$, $C_{ij}$, and $D_{ij}$, $(i,j=1,2)$ in Appendix III.

\section{Numerical Analysis}
%
\begin{figure}[t!]
\includegraphics[width=3in]{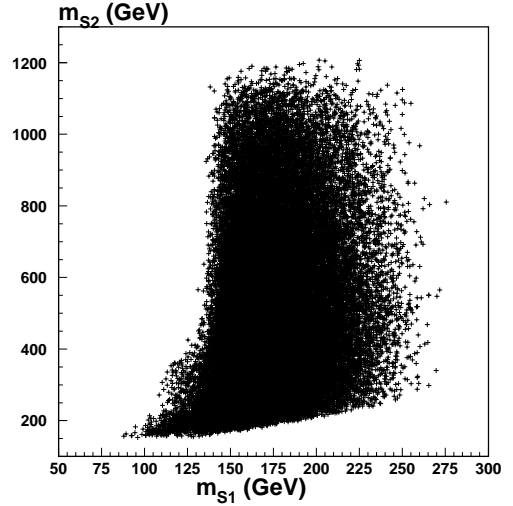}
\caption{The distribution of 50,000 points of ($m_{S_1}$, $m_{S_2}$), at the one-loop level.
The allowed ranges of the parameter values are
$0 < m_{du} \mbox{~(GeV)~} < 200$,
$2 < \tan \beta < 30$,
$150 < \mu \mbox{~(GeV)~} < 500$,
$500 < m_Q \mbox{~(GeV)~} < 700$,
$500 < A_t \mbox{~(GeV)~} < 700$,
$0 < k < 0.5$,
$300 < M_{\phi} \mbox{~(GeV)~} < 500$,
$300 < A_k \mbox{~(GeV)~} < 500$,
and $300 < \sqrt{b_{\phi}} \mbox{~(GeV)~} < 500$. }
\end{figure}

For numerical analysis, we simplify our formulae by assuming
that $k = h$, $M_Q = M_U$, $m_Q = m_T$, $m_1 = m_2 = m_3 = m_4$,
and $b_{\phi} = b_{\Phi}$.
We set $m_1 = m_2 = m_3 = m_4$ as $M_\phi$.
Note that these simplifications reduce the number of independent parameters,
but we are still left with a number of them: $m_{du}$ and $\tan \beta$ in $M^0$,
$\mu$, $m_Q = m_T$, and $A_t$ in $M^t$,
and $M_{\phi}$, $k$, and $b_{\phi}$ in $M^V$ and $M^{\tilde V}$.
We set the allowed ranges for these parameters as follows:
$0 < m_{du} \mbox{~(GeV)~} < 200$,
$2 < \tan \beta < 30$,
$150 < \mu \mbox{~(GeV)~} < 500$,
$500 < m_Q \mbox{~(GeV)~} < 700$,
$500 < A_t \mbox{~(GeV)~} < 700$,
$0 < k < 0.5$,
$300 < M_{\phi} \mbox{~(GeV)~} < 500$,
$300 < A_k \mbox{~(GeV)~} < 500$,
and $300 < \sqrt{b_{\phi}} \mbox{~(GeV)~} < 500$.
Here, the lower bound on the value of $\mu$ is set by the experimental constraint on the chargino system,
the upper bound on the Yukawa coupling coefficient of $k$ is set as 0.5 in order to
escape the Landau pole at a high energy scale.
The upper bound on the SUSY breaking soft masses on the scalar top quark sector are set as 700 GeV
because we do not want a large fine tuning.
For the top quark mass, we take 171 GeV and assume that it is smaller than the masses of the scalar top quarks.
It is obtained that the masses of the vector-like quarks lies in the
range of 300 GeV to 550 GeV, and the masses of the vector-like
scalar quarks in the range of 200 GeV to 910 GeV.

Now, let us study $G_{ZZS_i}$ ($i = 1, 2$), the coupling coefficients of the neutral scalar Higgs bosons to $ZZ$.
In the SM, the corresponding quantity is given as
\begin{equation}
    G_{ZZH} = g_2m_Z / \cos \theta_W
\end{equation}
where $\theta_W$ is the weak mixing angle, and $H$ is the neutral
scalar Higgs boson of the SM.
We introduce the normalized coupling coefficients
${\bar G}_{ZZS_i}$, which are defined as
$G_{ZZS_i} /G_{ZZH}$.
These normalized coupling coefficients satisfy a sum rule:
\begin{equation}
    \sum_{i=1}^2 {\bar G}^2_{ZZS_i} = 1 \ .
\end{equation}

For a given set of parameter values, we calculate the masses of the neutral scalar Higgs bosons at the one-loop
level, as well as their normalized coupling coefficients.
The LEP2 collaborations have established the model-independent upper bound on the squared coupling
coefficient of a neutral scalar boson to a pair of $Z$ bosons, as a function of the mass of the neutral scalar boson [48].
For given mass of $S_1$, we may compare ${\bar G}_{ZZS_1}$ with the LEP2 result.
If ${\bar G}_{ZZS_1}$ is larger than the upper bound set by the LEP2 result, we should increase the mass $S_1$
until ${\bar G}_{ZZS_1}$ becomes smaller than the LEP2 result.
In this way, the lower bound on the mass of $S_1$ can be established.
We examine 50,000 sets of parameter values, and select those sets that satisfy the LEP2 constraint.
Those parameter sets yield the masses of the vector-like quarks within the range of 300 to 550 GeV,
and the masses of the vector-like scalar quarks in the range of
$200 < m_{{\tilde V}_i} \mbox{~(GeV)~} < 910$ ($i=1-4$).

Our results are shown in Fig. 1, where we plot $(m_{S_1}, m_{S_2})$ for the selected parameter sets.
The lower bound on $m_{S_1}$ is determined by the LEP constraint.
One may note in Fig. 1 that the mass of $S_1$ at the one-loop level may be as large as 280 GeV.
This upper bound is reasonably larger than the previous result.
In the previous investigation, the mass of the lighter neutral scalar Higgs boson is predicted to be
less than 200 GeV [46].
Thus, the upper bound is considerably improved.
We attribute this improvement to the difference in approximations.
In Ref. [46], it is assumed that $M_Q, M_U \gg k v_u, h v_d$ and $b_{\Phi} = b_{\phi} = 0$.
Thus, the mass matrix of the vector-like scalar quark is decomposed into two $2 \times 2$ submatrices.
In our numerical analysis, these approximations are removed though some other simplifications are still used.
In other words, we explore a wider region in the parameter space.
We note that it is possible to examine the full parameter space of our model if no approximation is made in the
numerical analysis.
Nevertheless, $m_{S_1} \sim 280$ GeV can relieve the experimental pressure on our model in the sense that
it allows the future collider experiments a wider chance to discover the lightest scalar Higgs boson of our model.

\section{Conclusions}

We consider the Higgs sector in an extension of the MSSM with the vector-like quarks and scalar quarks.
A reasonable parameter region is set, and the neutral scalar Higgs boson masses are calculated using
the effective Higgs potential at the one-loop level,
where the radiative corrections due to the contributions from the loops of top quark, scalar top quarks,
as well as the vector-like quarks and vector-like scalar quarks, are taken into account.
We find that the contributions from the vector-like quark and scalar quark loops are significantly large,
if we calculate in detail, with fewer approximations, for a wider parameter space.
The mass of the lightest neutral scalar Higgs boson of our model at the one-loop level lies
within the range of 85 to 280 GeV, while being consistent with the LEP2 constraint.

\section*{Acknowledgments}

S. W. Ham thanks J. Y. Lee, P. Ko, Y. G. Kim for valuable comments.
He would like to acknowledge the support from KISTI under
"The Strategic Supercomputing Support Program (No. KSC-2008-S01-0011)"
with Dr. Kihyeon Cho as the technical supporter.
This research was supported by Basic Science Research Program
through the National Research Foundation of Korea (NRF) funded
by the Ministry of Education, Science and Technology (2009-0086961).

\section*{Appendix I}
\noindent The coefficients $A$, $B$ and $C$ in formulae (22) and (23) are expressed as in the following:
\begin{eqnarray*}
A =  {\cal A} + {\tilde {\cal A}} ,
\end{eqnarray*}
where
\begin{eqnarray*}
{\cal A} & = & -2 M_{Q\phi}^2-2 h^2 v_d^2 , \\
{\tilde {\cal A}} &=& {\cal A} (v_d \leftrightarrow v_u, h \leftrightarrow k, A_h \leftrightarrow A_k) ,
\end{eqnarray*}
and $M_{Q\phi}^2 = M_Q^2 + M_{\phi}^2$.
\begin{eqnarray*}
B  =  {\cal B} + {\tilde {\cal B}} , \nonumber
\end{eqnarray*}
where
\begin{eqnarray*}
{\cal B} &=& 3 M_Q^4+6 M_{\phi }^2 M_Q^2+6 h^2 v_d^2 M_Q^2+3 M_{\phi }^4+h^4 v_d^4-b_{\phi }^2 \cr
   & &-A_h^2 v_d^2+4 h^2 M_{\phi }^2 v_d^2+2 h^2 k^2 v_{du}^2-h^2 \mu ^2 v_u^2\cr
   & &-2 h k M_{\phi }^2 v_{du}+2 h \mu  A_h v_{du} ,\\
{\tilde {\cal B}} & = & {\cal B} (v_d \leftrightarrow v_u, h \leftrightarrow k, A_h \leftrightarrow A_k) ,
\end{eqnarray*}
where $v_{du}= (v^2 /2) {\sin}2\beta$.
\begin{eqnarray*}
C  =  {\cal C} + {\tilde {\cal C}} , \nonumber
\end{eqnarray*}
where
\begin{eqnarray*}
{\cal C} &=&-2 M_Q^6-6 M_{\phi }^2 M_Q^4-6 h^2 v_d^2 M_Q^4-6 M_{\phi }^4 M_Q^2 \cr
  & & -2 h^4 v_d^4 M_Q^2+2 A_h^2 v_d^2 M_Q^2-8 h^2 M_{\phi }^2 v_d^2 M_Q^2 \cr
   & & -4 h^2 k^2 v_{du}^2 M_Q^2+2 h^2 \mu ^2 v_u^2 M_Q^2 +4 h k M_{\phi }^2 v_{du} M_Q^2 \cr
   & & -4 h \mu  A_h v_{du} M_Q^2-2 M_{\phi }^6 +2 h^2 k^2 \mu ^2 v_d^4+2 b_{\phi }^2 M_{Q\phi}^2 \cr
   & & -2 h^2 M_{\phi }^4 v_d^2+2 h^2 b_{\phi }^2 v_d^2 +2 A_h^2 M_{\phi }^2 v_d^2+4 h k \mu  b_{\phi } M_{\phi } v_d^2 \cr
   & & -4 h A_h b_{\phi } M_{\phi } v_d^2+2 h^2 A_k^2 v_{du}^2 -2 h^2 k^2 M_{\phi }^2 v_{du}^2\cr
   & & -2 h^4 k^2 v_d^4 v_u^2 +2 h^2 \mu ^2 M_{\phi }^2 v_u^2+4 h k M_{\phi }^4 v_{du} \cr
   & & -4 h \mu  A_h M_{\phi }^2 v_{du}+4 h^2 \mu b_{\phi } M_{\phi } v_{du} -4 h A_k b_{\phi } M_{\phi } v_{du} \cr
   & & +4 h^3 k M_{\phi }^2 v_d^3 v_u -4 h^2 k \mu  A_k v_d^3 v_u  , \\
{\tilde {\cal C}} & = & {\cal C} (v_d \leftrightarrow v_u, h \leftrightarrow k, A_h \leftrightarrow A_k) .
\end{eqnarray*}
\section*{Appendix II}
\noindent
The coefficients $A_i$, $B_i$, $C_i$, and $D_i$ $(i = 1, 2)$ in
formulae (22) and (23) are expressed as in the following:
\begin{eqnarray*}
A_{1} &=&  -4 h^2 v_d, \\
A_{2} &=& A_1 (v_d \leftrightarrow v_u, h \leftrightarrow k, A_h \leftrightarrow A_k) , \\
B_{1} &=& 4 v_d^3 h^4+8 k^2 v_d v_u^2 h^2+12 M_Q^2 v_d h^2 \cr
    & &+8 M_{\phi }^2 v_d h^2-4 k M_{\phi }^2 v_u h+2 \mu  A_h v_u h \cr
    & &-2 k^2 \mu ^2 v_d-2 A_h^2 v_d+2 k \mu  A_k v_u, \\
B_{2} &=& B_1 (v_d \leftrightarrow v_u, h \leftrightarrow k, A_h \leftrightarrow A_k) , \\
C_{1} &=& -8 M_Q^2 v_d^3 h^4-8 k^2 v_d^3 v_u^2 h^4+12 k M_{\phi }^2 v_d^2 v_u h^3\cr
    & & -4 k^4 v_d v_u^4 h^2+8 k^2 \mu ^2 v_d^3 h^2+4 A_k^2 v_d v_u^2 h^2\cr
    & & -16 k^2 M_Q^2 v_d v_u^2 h^2-8 k^2 M_{\phi }^2 v_d v_u^2 h^2+4 b_{\phi }^2 v_d h^2\cr
    & & -12 M_Q^2 M_{Q\phi}^2 v_d h^2-4 M_{Q\phi}^2 M_{\phi }^2 v_d h^2 +4 k^3 M_{\phi }^2 v_u^3 h\cr
    & & -12 k \mu  A_k v_d^2 v_u h^2+4 \mu  b_{\phi } M_{\phi } v_u h^2 -4 k^2 \mu  A_h v_u^3 h\cr
    & & +8 k \mu  b_{\phi } M_{\phi } v_d h-8 A_h b_{\phi } M_{\phi } v_d h-4 \mu  A_h M_{Q\phi}^2 v_u h\cr
    & & +8 k M_{Q\phi}^2 M_{\phi }^2 v_u h-4 A_k b_{\phi } M_{\phi } v_u h +4 k^2 A_h^2 v_d v_u^2\cr
    & & +4 k^2 \mu^2   M_{Q\phi}^2 v_d+4 A_h^2 M_{Q\phi}^2 v_d +4 k^2 \mu  b_{\phi } M_{\phi} v_u\cr
    & & -4 k \mu  A_k M_{Q\phi}^2 v_u-4 k A_h b_{\phi } M_{\phi } v_u, \\
C_{2} &=& C_1 (v_d \leftrightarrow v_u, h \leftrightarrow k, A_h \leftrightarrow A_k) , \\
D_{1} &=& -6 h^4 k^2 \mu ^2 v_d^5+10 h^4 k \mu  A_k v_u v_d^4+4 h^4 M_Q^4 v_d^3 \cr
    & &+4 k^2 \mu ^2 A_h^2 v_d^3-8 h^2 k^2 \mu ^2 M_{Q\phi}^2 v_d^3+8 h^2 k \mu  A_h M_{\phi }^2 v_d^3\cr
    & &+8 h^4 k^2 M_Q^2 v_u^2 v_d^3-16 h^3 k \mu  b_{\phi } M_{\phi } v_d^3 +2 A_{hk}^2 v_{du} v_u\cr
    & &+12 h^2 k \mu A_k M_{Q\phi}^2 v_u v_d^2-6 h^3 k \mu ^2 M_{\phi }^2 v_u v_d^2\cr
    & &-12 h^3 k M_Q^2 M_{\phi }^2 v_u v_d^2+12 h k^2 \mu  A_h M_{\phi }^2 v_u v_d^2\cr
    & &-6 h^2 A_{hk} M_{\phi }^2 v_u v_d^2-6 h k^2 \mu ^3 A_h v_u v_d^2 -2 A_{hk} b_{\phi }^2 v_u\cr
    & &-12 h^2 k^2 \mu  b_{\phi } M_{\phi } v_u v_d^2+12 h^3 A_k b_{\phi } M_{\phi } v_u v_d^2\cr
    & &-4  h^4 A_k^2 v_{du} v_u v_d^2-2 k^2 \mu ^2 M_{Q\phi}^4 v_d-2 A_h^2 M_{Q\phi}^4 v_d\cr
    & &+4 h^2 M_Q^2 M_{Q\phi}^4 v_d-2 k^4 A_h^2 v_u^4 v_d+4 h^2 k^4 M_Q^2  v_u^4 v_d\cr
    & &+4 k \mu  A_h b_{\phi }^2 v_d-4 h^2 b_{\phi }^2 M_Q^2 v_d-8 h^2 b_{\phi }^2 M_{\phi }^2 v_d\cr
    & &-8 h k \mu  b_{\phi } M_{Q\phi}^2 M_{\phi } v_d+8 h A_h b_{\phi } M_{Q\phi}^2 M_{\phi } v_d\cr
    & &-6 k \mu  A_h^2 A_k v_{du} v_d+2 h k^4 \mu  A_h v_u^5-2 h \mu  A_h A_k^2 v_u^3\cr
    & &+4 h k^2 \mu  A_h M_{Q\phi}^2 v_u^3-2 h k^3 \mu^2 M_{\phi }^2 v_u^3 +4 h^4 k^4 v_{du}^3 v_u\cr
    & &-4 h k^3 M_Q^2 M_{\phi }^2 v_u^3-2 k^2 A_{hk} M_{\phi }^2 v_u^3 -4 h^2 k^2 b_{\phi }^2 v_{du} v_u \cr
    & &+4 h^2 k \mu  A_k M_{\phi }^2 v_u^3-2 h^2 k \mu ^3 A_k v_u^3-4 h^2 k^2 \mu  b_{\phi } M_{\phi } v_u^3\cr
    & &+4 k^3 A_h b_{\phi } M_{\phi } v_u^3+2 h \mu  A_h M_{Q\phi}^4 v_u+2 k \mu  A_k M_{Q\phi}^4 v_u\cr
    & &-4 h k M_{Q\phi}^4 M_{\phi }^2 v_u-4 h k b_{\phi }^2 M_{\phi }^2 v_u-12 h^3 k^3 M_{\phi }^2 v_{du}^2 v_u\cr
    & &-4 h^2 \mu  b_{\phi } M_{Q\phi}^2 M_{\phi } v_u-4 k^2 \mu  b_{\phi } M_{Q\phi}^2 M_{\phi } v_u\cr
    & &+4 k A_h b_{\phi } M_{Q\phi}^2 M_{\phi } v_u+4 h A_k b_{\phi } M_{Q\phi}^2 M_{\phi } v_u\cr
    & &+2 h^2 k^2 \mu ^4 v_{du} v_u+8 h^2 k^2 M_Q^4 v_{du} v_u -2 h k \mu ^2 b_{\phi }^2 v_u\cr
    & &+12 h^2 k^2 M_{\phi }^4 v_{du} v_u -4 k^2 A_h^2 M_{Q\phi}^2 v_{du} v_u\cr
    & &-4 h^2 A_k^2 M_{Q\phi}^2 v_{du} v_u-8 h^2 k^2 \mu ^2 M_{\phi }^2 v_{du} v_u\cr
    & &+8 h^2 k^2 M_Q^2 M_{\phi }^2 v_{du} v_u+4 k^3 \mu  A_h M_{\phi }^2 v_{du} v_u \cr
    & &-8 h k A_{hk} M_{\phi }^2 v_{du} v_u +4 h^3 \mu  A_k M_{\phi }^2 v_{du} v_u\cr
    & &+8 h k \mu ^2 A_{hk} v_{du} v_u +8 h k^2 A_h b_{\phi } M_{\phi } v_{du} v_u\cr
    & &+8 h^2 k A_k b_{\phi } M_{\phi } v_{du} v_u  ,  \\
D_{2} &=& D_1 (v_d \leftrightarrow v_u, h \leftrightarrow k, A_h \leftrightarrow A_k) ,
\end{eqnarray*}
\section*{Appendix III}
\noindent
The coefficients $A_{ij}$, $B_{ij}$, $C_{ij}$, and $D_{ij}$,
$(i,j=1,2)$ in formulae (22) and (23) are expressed as in the following:
\begin{eqnarray*}
A_{11} &=&-4 h^2,\\
A_{22} &=& A_{11} (v_d \leftrightarrow v_u, h \leftrightarrow k, A_h \leftrightarrow A_k) , \\
A_{12} &=&A_{21}=0, \\
B_{11} &=&12 v_d^2 h^4+12 M_Q^2 h^2+8 M_{\phi}^2 h^2 +8 k^2 v_u^2 h^2\cr
       & & -2 k^2 \mu^2-2 A_h^2,\\
B_{22} &=& B_{11} (v_d \leftrightarrow v_u, h \leftrightarrow k, A_h \leftrightarrow A_k) , \\
B_{12} &=&B_{21} = 8 h^2 v_{du} k^2-2 h M_{\phi }^2 k+2 h \mu  A_h,\\
C_{11} &=& -24 M_Q^2 v_d^2 h^4-24 k^2 v_{du}^2 h^4+24 k M_{\phi }^2 v_{du} h^3 \cr
    & & -12 M_Q^4 h^2-4 M_{\phi }^4 h^2-4 k^4 v_u^4 h^2+4 b_{\phi }^2 h^2 \cr
    & & -16 M_Q^2 M_{\phi }^2 h^2+24 k^2 \mu ^2 v_d^2 h^2+4 A_k^2 v_u^2 h^2 \cr
    & & -16 k^2 M_Q^2 v_u^2 h^2 -8 k^2 M_{\phi}^2 v_u^2 h^2 -24 k \mu  A_k v_{du} h^2 \cr
    & & +8 k \mu  b_{\phi}M_{\phi } h -8 A_h b_{\phi } M_{\phi } h+4 k^2 \mu^2 M_{Q\phi}^2 \cr
    & & +4 A_h^2 M_{Q\phi}^2+4 k^2 A_h^2  v_u^2 , \\
C_{22} &=& C_{11} (v_d \leftrightarrow v_u, h \leftrightarrow k, A_h \leftrightarrow A_k) , \\
C_{12} &=& C_{21} \cr
    &=& -16 k^2 v_d^3 v_u h^4+12 k M_{\phi }^2 v_d^2 h^3-12 k \mu  A_k v_d^2 h^2 \cr
    & &+4 \mu  b_{\phi } M_{\phi } h^2+8 A_k^2 v_{du} h^2-16 k^2 M_Q^2 v_{du} h^2 \cr
    & &-8 k^2 M_{\phi }^2 v_{du} h^2-4 \mu  A_h M_{Q\phi}^2 h+4 k M_{Q\phi}^2 M_{\phi }^2 h \cr
    & &-4 A_k b_{\phi } M_{\phi } h + {\tilde C}_{12} ,\\
D_{11} &=& -30 k^2 \mu ^2 v_d^4 h^4+12 k^4 v_d^2 v_u^4 h^4+12 M_Q^4 v_d^2 h^4\cr
    & &-12 A_k^2 v_{du}^2 h^4+24 k^2 M_Q^2 v_{du}^2 h^4+40 k \mu  A_k v_d^3 v_u h^4\cr
    & &-24 k^3 M_{\phi }^2 v_d v_u^3 h^3-48 k \mu  b_{\phi } M_{\phi } v_d^2 h^3 +2 A_{hk}^2 v_u^2\cr
    & &+4 \mu  A_k M_{\phi}^2 v_u^2 h^3-12 k \mu ^2 M_{\phi }^2 v_{du} h^3 -4 b_{\phi }^2 M_Q^2 h^2\cr
    & &-24 k M_Q^2 M_{\phi }^2 v_{du} h^3 +24 A_k b_{\phi } M_{\phi} v_{du} h^3 \cr
    & &+4 M_Q^4 M_{Q\phi}^2 h^2-8 b_{\phi }^2 M_{\phi }^2 h^2+4 M_Q^2 M_{Q\phi}^2 M_{\phi }^2 h^2\cr
    & &-24 k^2 \mu ^2 M_{Q\phi}^2 v_d^2 h^2+24 k \mu  A_h M_{\phi }^2 v_d^2 h^2 \cr
    & &+2 k^2 \mu ^4 v_u^2 h^2+8 k^2 M_Q^4 v_u^2 h^2+12 k^2 M_{\phi}^4 v_u^2 h^2\cr
    & &-4 k^2 b_{\phi }^2 v_u^2 h^2-4 A_k^2 M_{Q\phi}^2 v_u^2 h^2-8 k^2 \mu ^2 M_{\phi }^2 v_u^2 h^2\cr
    & &+8 k^2 M_Q^2 M_{\phi }^2 v_u^2 h^2+8 k A_k b_{\phi } M_{\phi } v_u^2 h^2 +4 k \mu  A_h b_{\phi}^2\cr
    & &+24 k \mu  A_k M_{Q\phi}^2 v_{du} h^2 -12 A_{hk} M_{\phi }^2 v_{du} h^2 \cr
    & &-24 k^2 \mu b_{\phi } M_{\phi } v_{du} h^2-8 k A_{hk} M_{\phi}^2 v_u^2 h -2 A_h^2 M_{Q\phi}^4\cr
    & &+8 k \mu ^2 A_{hk} v_u^2 h+8 k^2 A_h b_{\phi } M_{\phi } v_u^2 h +12 k^2 \mu ^2 A_h^2 v_d^2\cr
    & &-8 k \mu  b_{\phi } M_{Q\phi}^2 M_{\phi } h -12 k^2 \mu ^3 A_h v_{du} h -2 k^4 A_h^2 v_u^4 \cr
    & &-4 k^2 A_h^2 M_{Q\phi}^2 v_u^2  +4 k^4 M_Q^2 v_u^4 h^2 -2 k^2 \mu ^2 M_{Q\phi}^4\cr
    & &+8 A_h b_{\phi } M_{Q\phi}^2 M_{\phi } h+24 k^2 \mu  A_h M_{\phi }^2 v_{du} h \cr
    & &+4 k^3 \mu  A_h M_{\phi }^2 v_u^2 -12 k \mu  A_h^2 A_k v_{du},\\
D_{22} &=& D_{11} (v_d \leftrightarrow v_u, h \leftrightarrow k, A_h \leftrightarrow A_k) , \\
 D_{12} &=& D_{21} \cr
    &=& 8 h^4 v_u^3 k^4-8 A_h^2 v_d v_u^3 k^4+16 h^2 M_Q^2 v_d v_u^3 k^4 \cr
    & &-18 h^3 M_{\phi }^2 v_{du}^2 k^3+12 h \mu  A_h M_{\phi }^2 v_d^2 k^2 \cr
    & &-12 h^2 \mu  b_{\phi } M_{\phi} v_d^2 k^2 +2 h^2 \mu ^4 v_{du} k^2 +8 h^2 M_Q^4 v_{du} k^2\cr
    & &+12 h^2 M_{\phi }^4 v_{du} k^2 -4 h^2 b_{\phi }^2 v_{du} k^2-8 h^2 \mu ^2 M_{\phi }^2 v_{du} k^2\cr
    & &+8 h^2 M_Q^2 M_{\phi }^2 v_{du} k^2+10 h^4 \mu  A_k v_d^4 k -h \mu ^2 b_{\phi }^2 k\cr
    & &-2 h M_{Q\phi}^4 M_{\phi }^2 k-2 h b_{\phi }^2 M_{\phi }^2 k +12 h^2 \mu A_k M_{Q\phi}^2 v_d^2 k\cr
    & &-6 h^3 \mu ^2 M_{\phi }^2 v_d^2 k -12 h^3 M_Q^2 M_{\phi }^2 v_d^2 k-6 \mu  A_h^2 A_k v_d^2 k\cr
    & &-8 h A_{hk} M_{\phi }^2 v_{du} k +8 h \mu ^2 A_{hk} v_{du} k +2 A_{hk}^2 v_{du} \cr
    & &+16 h^2 A_k b_{\phi } M_{\phi } v_{du} k +2 h \mu  A_h M_{Q\phi}^4-A_{hk} b_{\phi }^2 \cr
    & &-6 h^2 A_{hk} M_{\phi }^2 v_d^2 +12 h^3 A_k b_{\phi } M_{\phi } v_d^2 -6 h \mu ^3 A_h v_d^2 k^2\cr
    & &-4 h^2 \mu  b_{\phi } M_{Q\phi}^2 M_{\phi }+4 h A_k b_{\phi } M_{Q\phi}^2 M_{\phi} \cr
    & &-8 h^2 A_k^2 M_{Q\phi}^2 v_{du}+8 h^3 \mu  A_k M_{\phi }^2 v_{du} +{\tilde D}_{12},
\end{eqnarray*}
where $A_{hk}=A_h A_k$.

\end{document}